\documentstyle[12pt]{article}

\begin{document}

\title{ELECTRONS ON ROTATIONALLY SYMMETRIC NANOPARTICLES UNDER A STRONG MAGNETIC
FIELD}
\author{P.Malits$^{1,2}$, A. Kaplunovsky$^{1,2}$, I. D. Vagner$^{1,2}$ and P. Wyder$%
^{1}$ \\
$^{1}$Grenoble High Magnetic Field Laboratory \\
Max-Planck-Institut f\"ur Festk\"orperforschung and \\
Centre Nationale de la Recherche Scientifique, \\
BP 166, 38042 Grenoble Cedex 09, France.\\
$^2$Physics and Engineering Research Institute at Ruppin,\\
Emek Hefer, 40250\\
Israel.}
\maketitle

\begin{abstract}
The energy spectrum of an electron confined to an arbitrary surface of
revolution in an external magnetic field, parallel to the symmetry axis, is
studied analitycally and numerically. The problem is reduced via conformal
mapping to one on the surface of a sphere. The case of a spheroid is
considered in details, and the dependence on parameters is discussed. In the
high magnetic field limit a regular structure in the energy spectrum,
resembling the Landau levels, is obtained. Level statistics is discussed.

\vfill
\end{abstract}

\ 

Recent technological progress in fabrication of semiconducting and metallic
nanostructures opened a vast field of research of their electron properties 
\cite{PWM81, RvL91}. Traditionally application of high magnetic fields is an
extremely powerful method for experimental studies of electronic properties
in solids. Detailed theoretical study of the electron spectrum of the
nano-structures under strong magnetic fields is therefore of primary
importance for the future progress in this field . Initially the solutions
for an electron in confined plane geometries, like a disc, ring, cylinder
and oval shape stadium \cite{BIL83,WEFZS91,NT88} were proposed. As it was
shown in \cite{NT88} , these models are relevant to the notion of the chaos
in the level statistics and related thermodynamics of such systems. Among
already studied three dimensional systems are electrons on simple surfaces
as nanotubes \cite{NT88} and spheres \cite{KVS92,Aoki93}. A challenging
problem is an adequate quantum mechanical description of noninteracting
electrons on a nanoparticle of an arbitrary shape.

Here we consider a single electron confined to the surface of revolution
placed in an axial uniform magnetic field . Our goal is to treat the general
case of the arbitrary shaped surface of revolution $r=f(z)$ ( $(r,\varphi
,z) $ are cylindrical coordinates) and to investigate influence of its
geometrical characteristics upon quantum- mechanical spectrum. Further we
suppose the surface to be smooth , closed and crossing $z$-axes only in two
points. The uniform magnetic field $B$ is defined to point in the $z$%
-direction .

The problem is described by the Hamiltonian

\begin{equation}
{\cal H}=\frac 1{2m}\left[ ih \nabla -{\bf A}\right] ^2+V{ ,}
\label{ham11}
\end{equation}
where , for simplicity ,we ignore spin dependent terms . ${\bf A=}%
B(-y,x,0)/2 $ is the symmetric gauge , $(x,y)$ are Cartesian coordinates.
This is leading to the Schrodinger equation on the surface $r=f(z)$

\begin{eqnarray}
(\Delta +2iB_1\frac \partial {\partial \varphi }-B_1^2r^2-V_1)\psi
&=&-E_1\psi { ,}  \label{Sch1}
\end{eqnarray}
where: $E_1=\frac{2m}{h ^2}E$ , $B_1=\frac{eB}{2ch }$ $,$ $V_1=\frac{%
2m}{h ^2}V$ .

We introduce new orthogonal coordinates by $z+ir=F(u+iv)$ , where the
function $F(u+iv)$ maps conformally the domain of the $(u,v)$-plane
containing the unit circle onto the domain of the $(r,z)$-plane containing
the closed curve $r=\pm f(z).$ This curve is the image of the circle $%
u^2+v^2=1$ with the arc $0\leq \theta $ $\leq \pi $ $\left( u+iv=R\exp
i\theta \right) $corresponding to $r\geq 0$. Since conformal mapping
conserves a normal to the surface, it enables us to write the Eq.\ref{Sch1}
on the surface $R=1$ neglecting derivatives in $R.$ Thus , the three
dimensional Schrodinger operator has been reduced to a two-dimensional
operator in $(\theta ,\varphi )$-variables .

Due to conservation of the $z$-component of the angular momentum, the cyclic
coordinate $\varphi $ can be separated in the Fourier series development

\begin{equation}
\psi \left( \theta ,\varphi \right) =\sum_{m=-\infty }^{+\infty }\psi
_m\left( \theta \right) \exp \left( im\varphi \right) { . }  \nonumber
\end{equation}

Further simplification $x=\cos $ $\theta $ results in the ordinary
differential equation of the second order

\begin{center}
\begin{equation}
\left( 1-x^2\right) \frac{d^2\psi _m}{dx^2}-G_1\left( x\right) \frac{d\psi _m%
}{dx}+G_0\left( x\right) \psi _m=0{ , }\left| x\right| \leq 1{ ,}
\label{Eq1}
\end{equation}
\[
\left| \psi _m\left( \pm 1\right) \right| <\infty { .} 
\]
\end{center}

Here : $G_1\left( x\right) =x-\left( 1-x^2\right) \rho ^{\prime }\left(
x\right) \rho ^{-1}\left( x\right) $, $\rho _0\rho \left( x\right) =Im$ $%
F\left( x+i\sqrt{1-x^2}\right) $,

$\rho _0=\max Im$ $F\left( \exp \left( i\theta \right) \right) ,$ $G_0$ $%
\left( x\right) =\Phi \left( x\right) \left[ \lambda -\widetilde{B}^2\rho
^2\left( x\right) -m^2\rho ^{-2}\left( x\right) \right] \rho _0^{-2},$

$\Phi \left( x\right) =\left| F^{\prime }\left( x+i\sqrt{1-x^2}\right)
\right| ^2$, $\lambda =\widetilde{E}-2\widetilde{B}m$ , $\widetilde{E}%
=(E_1-V_{1)}\rho _0^2$, $\widetilde{B}=B_1\rho _0^2$ .

A low field $\left( \widetilde{B}\ll 1\right) $asymptotics of the spectrum
and eigenfunctions may be found in the traditional way by the pertubation
method . It is much more difficult to suggest some general approach to
indicate a high field $\left( \widetilde{B}\gg 1\right) $asymptotics. This
is governed by coefficients of the Eq.\ref{Eq1} or , in other words , by the
surface shape. Some possible shapes and corresponding conformal mappings $%
\left( \varsigma =u+iv\right) $ are shown on Fig 1-3 and \cite
{mkvw98} ,where conformal
mappings are pointed out in the brackets.

Below we consider closely a spheroidal surface whose equation is 
\begin{equation}
\frac{z^2}{a^2}+\frac{r^2}{b^2}=1{ }.  \label{el1}
\end{equation}

Conformal mapping: $z+ir=\frac{a-b}{2\varpi }+\left( a+b\right) \frac \varpi
2$ $,$ $\varpi =R\exp \left( i\theta \right) $ $,$ is an one-to-one mapping
of the unit circle $R=1$ onto this ellipse of the $\left( r,z\right) $-plane

The Eq.\ref{Eq1} can be written in the form

\begin{equation}
\frac d{dx}\left( 1-x^2\right) \frac{d\psi _m}{dx}+\left[ \lambda -%
\widetilde{B}^2\left( 1-x^2\right) -\frac{m^2}{1-x^2}\right] \left[ \xi
\left( 1-x^2\right) +1\right] \psi _m=0  \label{eqel1}
\end{equation}

\[
\left| x\right| \leq 1{ , }\left| \psi _m\left( \pm 1\right) \right|
<\infty { ; }\rho _0=b{ , }\xi =a^2b^{-2}-1{ .} 
\]
This problem has an infinite discrete spectrum $\lambda _{lm}$ . Its
eigenfunctions $\psi _{lm}\left( x\right) $ have $l$ zeroes in the interval $%
\left( -1,1\right) $ . One can see, that if $l$ is even (odd) integer, then
these functions are even (odd) .

It can be shown that all eigenvalues $\lambda _{lm}$ are positive . They are
large as one of the conditions : 1) $l\gg 1$ , 2) $m\gg 1$ , 3) $\widetilde{B%
}\gg 1$ , is fulfilled . Below we are pointing out leading terms of the
corresponding asymptotics .

As $l\gg 1$ , the spectrum can be obtained with the method of the paper \cite
{a3}. Particularly , the leading term is given by

\begin{equation}
\widetilde{E}_{lm}=\frac{\pi ^2\left( 2l+2\left| m\right| +1\right) ^2}{%
16\left( 1+\xi \right) {\bf E}^2\left( \sqrt{\frac \xi {1+\xi }}\right) }%
+2Bm+O\left( \frac 1l\right) ,  \label{lam1}
\end{equation}
where ${\bf E}\left( x\right) $ is a complete elliptic integral of the
second kind .

As $m\gg 1$ , the asymptotic expansion may be found with a stretched
variable $\lambda _{lm}=\widetilde{E}_{lm}-2\widetilde{B}m=m^2+(2l+1)\frac{%
\sqrt{\xi }}{1+\xi }\left| m\right| +O\left( 1\right) $ .Eigenfunctions are
expressed by Hermite polynomials $\psi _{lm}\left( x\right) =\exp \left( -x^2%
\sqrt{\lambda _{lm}}\right) H\left( x\lambda _{lm}^{\frac 14}\right)
+O\left( m^{-2}\right) $ , $x\in \left( -\varepsilon ,\varepsilon \right) .$

In the high field limit ( $\widetilde{B}\gg 1$ ) the spectrum is given by an
asymptotic formula:

\begin{equation}
\widetilde{E}_{lm}=2N\widetilde{B}-\frac 12N\left( N-2m\right) \left( \xi
+1\right) +\frac 12\left( \xi -1\right) +O\left( \frac 1{\widetilde{B}%
}\right) ,N={ }l+\cos ^2\frac{\pi l}2+\left| m\right| +m{.}
\label{lalev1}
\end{equation}

The corresponding asymptotic expansion of the eigenfunctions is expressed by
Laguere polynomials

\[
\psi _{lm}\left( x\right) =\left( \frac x{\left| x\right| }\right) ^l\left(
1-x^2\right) ^{\frac{\left| m\right| }2}\exp \left( -\frac 12\left(
1-x^2\right) \widetilde{B}\right) L_n^{\left| m\right| }\left( \left(
1-x^2\right) \widetilde{B}\right) +O\left( \frac 1{\widetilde{B}}\right) 
\]

\[
n=\frac 12\left( l-\sin ^2\frac{\pi l}2\right) ,x\notin \left( -\varepsilon
,\varepsilon \right) 
\]

Hence it appears that bunches of the energy levels resembling the Landau
levels are formed in the high field limit. Every bunch consists of the
parallel equidistant levels with the same number $N.$ Their leading term is
irrespective of a spheroidal geometry and coincide with the spectrum for the
plane. The energy level corresponds to two quasi-degenerated bound states
labeled $(2k,m)$ and $\left( 2k+1,m\right) .$

A disk of the radius $\rho _0$ is a limiting case of a strongly flattened
spheroidal shell ($\xi =-1$). In this limit , the values of the
eigenfunctions on both sides of the disk $\left( x>0{ and }x<0\right) $
are added and according to the above formula the antisymmetric
eigenfunctions are cancelled out . Since $\rho _0^2\left( 1-x^2\right) =r^2$
, we obtain

\[
E_1-V_1=2B_1\left( 2n+\left| m\right| +m+1\right) +\frac 1{\rho
_0^2}+O\left( \frac 1{\widetilde{B}\rho _0^2}\right) { ,} 
\]

\begin{equation}  \label{lan!}
\end{equation}
\[
\psi _{nm}\left( r\right) =r^{\left| m\right| }\exp \left( -\frac
12r^2B_1\right) L_n^{\left| m\right| }\left( r^2B_1\right) +O\left( \frac 1{%
\widetilde{B}}\right) { , }r<\rho _0-\varepsilon { .} 
\]

These relationships turns into the well known Landau solution as $\rho
_0=\infty $. In the high magnetic field the disc (circle billiard) spectrum
coalesces into the straight lines nearly the same that in the classical
Landau problem . This confirms results of the numerical calculations by
K.Nakamura and H.Thomas \cite{NT88}.

The algorithm of the calculation of the spectrum in a general case is given in 
\cite{mv99}.
Results of the calculations are represented on the Fig 4-5 \cite{mkvw98} for

$\xi =-0.7;$ $|m|=0,...,4;$ $l=0,...,7$ (Fig 4.) and

$\xi =3.0;$ $|m|=0,...,4;$ $l=0,...,7$ (Fig 5.)

We observe the behavior of the spectrum that was derived above analytically.
As $\xi $ or $N$ increase, the distance between lines of the same bunch( the
splitted Landau level) enlarges. Irregular crossings intensify as well.
These crossings arise from dropping and intermingling lines of the different
bunches.

Examples of the wave functions for the various parameters are shown on the
Fig 6, where

$\widetilde{B}=3;$ $m=3;$ $l=0;$ $\xi =0.7;0;1;3$

We may suggest that the electron motion on an arbitrary shaped convex
surface has the spectrum behavior similar to the spectrum behavior of the
electron on the spheroid. In the high field
region the spectrum behavior is predetermined by the surface flatness in the
vicinity of the poles (where the electron is trapped), and energy levels
constitute bunches of the straight lines that coalesces into the Landau
levels as the surface is strongly flat $\left( \xi _{0}\ll 1\right) .$


\begin{thebibliography}{99}
\bibitem{PWM81}  J.A.A.J. Perenboom, P. Wyder and F. Meier, Physics Reports,
(1981)

\bibitem{RvL91}  J.M. van Ruitenbeck and D.A. van Leuwen, Phys. Rev. Lett., 
{\bf 67}, 640 (1991).

\bibitem{BIL83}  M. B\"uttiker, Y. Imry and R. Landauer, Phys. Lett., {\bf %
96A}, 365 (1983).

\bibitem{WEFZS91}  D. Wohleben et al., Phys. Rev. Lett., {\bf 66}, 3191
(1991).

\bibitem{NT88}  K. Nakamura and H. Thomas, Phys. Rev. Lett. {\bf 61, }247
(1988).

\bibitem{KVS92}  Ju H. Kim, I.D. Vagner and B. Sundaram, Phys. Rev. B {\bf 46%
}, 9501 (1992).

\bibitem{Aoki93}  H.Aoki and H. Suezava, Phys. Rev. A{\bf 46}, R1163 (1992).

\bibitem{mkvw98}  P. Malits, A. Kaplunovsky, I.D. Vagner, and P. Wyder, High-Mag Theory Journal, http://www.magniel.com/hmtj/papers/mkvw98.html
(1998)

\bibitem{a3}  M.V.Fedorjuk, Diff.Equations, v.18, 2166-2173 (1982); v.19,
278-286 (1983).

\bibitem{mv99}  P. Malits and I.D. Vagner, J. Phys. A: Math. Gen., 32, 1507 (1999).



\end{thebibliography}
\end{document}